\documentclass[twocolumn,superscriptaddress,showpacs,preprintnumbers,prl]{revtex4}

\usepackage{graphicx}
\usepackage{dcolumn}
\usepackage{bm}
\usepackage{amsmath}
\usepackage{hyperref}
\usepackage{color}

\newcommand{\caco}{$\mathrm{Ca_3Co_2O_6 }$}

\begin{document}

\title{ Resonant X-ray scattering investigation of the multipolar ordering  
in \caco}

\author{A. Bombardi}
\affiliation{Diamond Light Source Ltd., Rutherford Appleton
Laboratory, Chilton-Didcot OX11-0QX, UK }
\author{C. Mazzoli}
\affiliation{European Synchrotron Radiation Facility, BP 220,
38043 Grenoble Cedex 9, France}
\author{S. Agrestini}
\affiliation{Department of Physics, University of Warwick, Coventry, CV4 7AL, UK}
\author{M.R. Lees}
\affiliation{Department of Physics, University of Warwick, Coventry, CV4 7AL, UK}

\date{\today}

\begin{abstract}
We have performed a resonant x-ray scattering (RXS) study near the Co K edge on a single crystal of \caco. 
In the magnetically ordered phase a new class of weak reflections appears at the magnetic propagation vector 
$\vec{\tau} \simeq (\frac{1}{3},\frac{1}{3},\frac{1}{3})$. These new reflections allow direct access to the dipolar-quadrupolar $E_1E_2$ scattering channel. The theoretical possibility of observing isolated $E_1E_2$ electromagnetic multipoles has attracted a lot of interest in the recent years. Unfortunately in many system of interest, parity even and parity odd tensor contributions occur at the same positions in reciprocal space. We demonstrate that in \caco\ it is possible to completely separate the parity even from the parity odd terms. The possibility of observing such terms even in globally centrosymmetric systems using RXS has been investigated theoretically; \caco\ allows a symmetry based separation of this contribution.        

\end{abstract}

\pacs{75.25.+z, 75.50.Ee, 78.70.Ck}

\maketitle


\caco\ has until very recently been described in terms of ferromagnetic chains coupled antiferromagnetically 
on a triangular lattice. The discovery of a small incommensurability~\cite{Agrestini08} in the magnetic propagation vector along the $c$ axis has been crucial in leading to a reevaluation of the role of the local exchange integrals in the description of the system. 
The system, shown in Fig. 1, consist of chains made up of alternating distorted octahedra and trigonal CoO$_6$ prisms sharing faces, running along the hexagonal $c$ axis and arranged in a triangular pattern within the $ab$ plane~\cite{Fjellvag96}. 
The different local environments leave the Co$^{3+}$ ions on the octahedral site (CoI) in a low-spin (S = 0) state, and those on trigonal prism (CoII)  sites in the high-spin (S = 2) state~\cite{Sampa04,Burnus06}. The local anisotropy of the trigonal prism forces the magnetic moments to point along the $c$ axis as confirmed by a number of experimental results.  \caco\ is usually described in the hexagonal setting of the \emph{R$\overline{3}$c} space group, as this representation allows one to immediately identify the triangular arrangement of the CoO$_6$ chains within the \emph{ab} planes. However, as this setting is non primitive, it makes both the description of the magnetic structure and the analysis of the symmetry of the tensors more difficult. For these reasons, in this letter the rhombohedral axes are used throughout. 
In the rhombohedral setting, the unit-cell dimensions are a=b=c=6.274 \AA ~ and $\alpha=\beta=\gamma=92.53^o$. 
Compared with other transition metal ions, the Co$^{3+}$ ion carries a large orbital moment that in \caco\ reaches 1.7 $\mu_B$ with $L/S \simeq 0.94$~\cite{Burnus06}. 
This large orbital contribution is crucial to the observation of small effects such as the $E_1E_2$ signal described in the following. 

\begin{figure}
\includegraphics[width=\columnwidth]{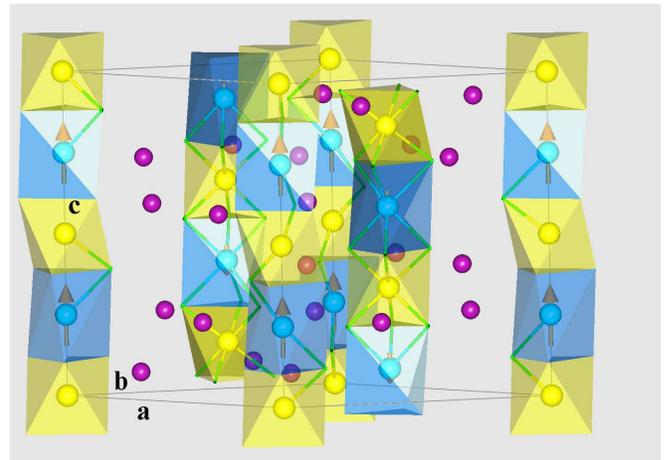}
\vspace{-0.5cm}
\caption{\label{fig:Ca3Co2O6} (Color on line) Schematic of the unit cell of \caco. The trigonal prisms are shown in light blue (dark grey) and the octahedra 
in yellow (light grey). The magnetic moments are also drawn.}
\end{figure}  

In this letter we report the results of a resonant x-ray investigation on \caco\ performed near the Co K-edge. 
In the magnetically ordered state, in addition to the principal magnetic reflections, we observe a second class of reflections that also appear at the magnetic propagation vector $\vec{\tau} \simeq (\frac{1}{3},\frac{1}{3},\frac{1}{3})$ but at positions where the magnetic structure factor is zero.  These reflections are characterized by a completely different photon energy spectra and azimuthal dependence. We first demonstrate that unlike the principal magnetic reflections, they cannot be described within the dipolar-dipolar ($E_1E_1$) and quadrupolar-quadrupolar ($E_2E_2$) approximation. We then provide a fit based on a magnetic contribution coming from the dipolar-quadrupolar ($E_1E_2$) interference. The occurrence of higher rank scattering in the space group \emph{R$\overline{3}$c} was treated by a number of authors ~\cite{Dmitrienko,Carra94,DiMatteo03,Lovesey05}. The case of V$_2$O$_3$~\cite{Paolasini01}, where a new family of reflections was discovered, looks similar to our situation. A very important difference between \caco\ and V$_2$O$_3$ is the presence of a first order phase transition in V$_2$O$_3$. In fact, there is no symmetry left at the V site in the low temperature phase of V$_2$O$_3$. For \caco, no detectable structural changes are observed as the system is cooled, even though the magnetic symmetry of the system is lower than the structural symmetry. 
Therefore, we preserve all the symmetry of the space group when we consider the charge distribution contribution to the scattering. 


Single crystals of \caco\ were grown by a flux method. Two crystals with different orientations were used for the RXS experiments. The first crystal was $5~\times~2~\times~1$~mm$^3$ with the largest natural face perpendicular to the $(1\,0\,\overline{1})$ reflection. The second crystal was polished to have a diffraction face perpendicular to the $(1,1,1)$ direction. The RXS experiments were performed at the magnetic scattering beamline ID20~\cite{Paolasini07} at the ESRF (Grenoble). The beamline optics were optimized at 7.7 keV, close to the Co K-edge. 
The sample was mounted in a displex cryostat.  The diffractometer was operated in the vertical plane scattering mode with an azimuth set-up to allow for a sample rotation about the scattering vector. Hence, the natural polarization of the incident beam was perpendicular to the scattering plane ($\sigma$). The integrated intensity of the reflections was measured using a photon counting avalanche photodiode detector. The polarization of the reflected beam was linearly analyzed by rotating the scattering plane of a highly oriented $\langle 00L \rangle$ pyrolitic graphite plate.

In the resonant regime, the elastic scattering amplitude, $A(\vec{Q},\omega)$, can be written in terms of the tensorial atomic scattering factor $f_j(\omega)$ with the index $j$ running over all the atoms contributing to the scattering in the unit cell:

\begin{equation}
A(\vec{Q},\omega)= \sum_j f_j(\omega) e^{i \vec{Q} \cdot \vec{r}_j}
\end{equation}

\noindent where $\hbar \vec{Q}$ is the momentum transfer in the scattering process and  $\hbar \omega$ is the energy of the 
incoming and outgoing photons. The two magnetic Co ions in the unit cell lie on the trigonal axis at the special positions $Co_1=(\frac{1}{4},\frac{1}{4},\frac{1}{4})$ and $Co_2=(\frac{3}{4},\frac{3}{4},\frac{3}{4})$, related by the inversion centre $\hat{I}$ at $(0,0,0)$. We note that due to the small incommensurability of the magnetic propagation vector, the value of the moment at site two is not identical to the moment at site one, therefore the inversion symmetry is locally broken. However, as the difference between the moment on the two ions is small, we neglect this difference and we keep this symmetry operation. As the magnetic structure of the system is incommensurate, it is still possible to find one inversion point in the magnetized crystal. Therefore globally the system cannot be ferroelectric.  

The non-magnetic Co ions that may contribute to a non-magnetic resonant process lie on the inversion center ($Co^{nm}_1=(0,0,0)$, $Co^{nm}_2=(\frac{1}{2},\frac{1}{2},\frac{1}{2})$) and they are related by an axis {\emph 2}. 
In the following, we neglect these ions as an analysis of the tensors contributing to the scattering shows that the only tensors that may contribute to the scattering at the reflection positions ($\hat{F}^4_3+\hat{F}^4_{-3}$) produce an azimuthal dependence that does not match with the experimental observations.

The scattering amplitude due to the magnetic Co ions is given by:

\begin{equation}
A(\vec{Q},\omega)= e^{i \vec{Q} \cdot (\frac{1}{4},\frac{1}{4},\frac{1}{4})} \left( f_1+\hat{I}f_1 e^{i  \vec{Q} \cdot (\frac{1}{2},\frac{1}{2},\frac{1}{2})}\right).
\label{eq:1}
\end{equation}

\noindent The tensorial atomic scattering factor $f_j(\omega)$ is a complex quantity whose explicit form is given in many papers.  
Following Di Matteo et al.~\cite{DiMatteo03} it is useful to write 

\begin{equation}
f_j(\omega)= \sum_{p,q,}(-)^q T^p_q F^p_{-q}(j;\omega),
\end{equation}

\noindent where the irreducible tensors $T^p_q$ describe the scattering process and are a function of 
the incident and outgoing polarizations and wave vectors, while the irreducible tensors of the same rank $F^p_{-q}(j;\omega)$ describe the multipolar properties of the system. It is useful to evaluate the tensors $T^p_q$ and $F^p_{-q}(j;\omega)$ in two different reference frames and to rotate $F^p_{-q}(j;\omega)$ to bring the tensors into the same reference frame. $T^p_q$ are calculated in the reference frame given in Ref.~\onlinecite{Lovesey05}, whereas the crystal reference frame has been chosen with the $\hat{z}$ axes parallel to the $(111)$ direction, the axis $\hat{y}$ parallel to the $(\overline{1}01)$ direction and the axis $\hat{x}$ parallel to $\hat{y} \times \hat{z}$.  
 
The $F^p_{-q}$ tensors allowed by symmetry need to be invariant under the point group of the system.
The non-magnetic point group symmetry is \emph{32}. We note that maintaining this point group would lead to the absence of a $E_1E_1$ magnetic contribution as expected since \emph{32} is not an admissible magnetic point group. 

The admissible magnetic point group is \emph{32'}. The axis \emph{3} is maintained and since the proper rotations act on spins in the same way as on polar vectors, we need to reverse the action of the axis \emph{2} to maintain the spin invariance under the point group.     
The tensors that survive to the axis \emph{3} are as follows: (i) in the $E_1E_1$ channel:  $F^{0}_0,\tilde{F}^{1}_0,F^{2}_0$; 
(ii) in the $E_1E_2$ channel:  $\tilde{F}^{1}_0,\tilde{F}^{2}_0,\tilde{F}^{3}_0,\tilde{F}^{3}_3 \pm \tilde{F}^{3}_{-3}$, and $F^{1}_0,F^{2}_0,F^{3}_0,F^{3}_3 \pm F^{3}_{-3}$; (iii) in the $E_2E_2$ channel: $F^{0}_0,\tilde{F}^{1}_0,F^{2}_0,\tilde{F}^{3}_0,\tilde{F}^{3}_3 \pm \tilde{F}^{3}_{-3},F^{4}_0,F^{4}_3 \pm F^{4}_{-3}$. 
The tilde indicates the time reversal odd (i.e. magnetic) tensors. 

A further reduction in the number of the tensors contributing to the scattering is obtained by applying the axis \emph{2} on the non-magnetic tensors and \emph{2'} if we consider the magnetic quantities. We are left with:
(i) $E_1E_1$: $F^{0}_0,\tilde{F}^{1}_0,F^{2}_0$; (ii) $E_1E_2$: $\tilde{F}^{1}_0, \tilde{F}^{3}_0,\tilde{F}^{3}_3 - \tilde{F}^{3}_{-3}$ and $F^{2}_0, F^{3}_3 + F^{3}_{-3}$; (iii) $E_2E_2$:  $F^{0}_0, F^{2}_0,\tilde{F}^{3}_3 + \tilde{F}^{3}_{-3},F^{4}_0,F^{4}_3 - F^{4}_{-3}$. It is easy to see $\tilde{F}^{1}_0$ is retained and that it corresponds to the magnetic moment 
along the trigonal axis.
  
\begin{figure}
\includegraphics[width=\columnwidth]{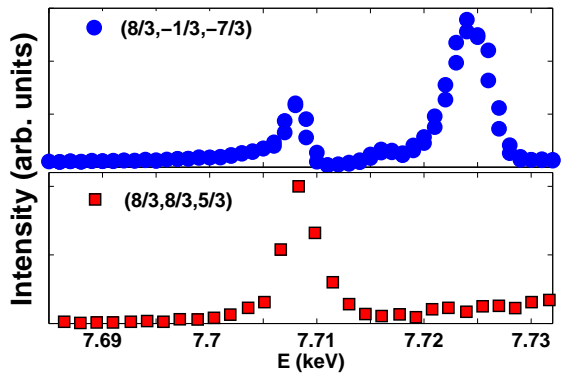}
\vspace{-0.7cm}
\caption{\label{fig:escan017_331} (Color on line) The intensity versus photon-energy dependence around the Co K-absorption edge of an $E_1E_1$ (top) and an $E_1E_2$ (bottom) magnetic reflection. The data were collected in the $\sigma \pi$ channel.}
\end{figure}

There are still a considerable number of contributions. An effective way to simplify the problem is to calculate the structure factor in Eq. \ref{eq:1} for the reflections $\vec{Q}= (H,K,L) \pm (\tau,\tau,\tau)$, with $\tau=\frac{1}{3}$ for the magnetic reflections and $\tau =0$ for the structural reflections. 
In the magnetic case we start with $H+K+L=$ even, then $H+K+L \pm 3\tau=$ odd and  
$A(hkl \pm \vec{\tau},\omega) = \pm i (F^p_{q,1}-(-1)^P F^p_{q,1})$, where $P$ is the parity of the tensor. 
Only the parity odd tensor, $E_1E_2$ processes, will contribute to the scattering process. Considering the forbidden reflections $H+K+L=$ odd, then $H+K+L \pm 3\tau=$ even and $A(hkl \pm \vec{\tau},\omega) = \pm (F^p_{q,1}+(-1)^P F^p_{q,1})$, we see that only the $E_1E_1$ and the $E_2E_2$ processes will contribute to the scattering process at these positions.


In Fig. \ref{fig:escan017_331} we report the photon-energy dependence around the Co K-absorption edge of the $(\frac{8}{3},\frac{8}{3},\frac{5}{3})$  and the $(\frac{8}{3},\frac{\overline{1}}{3},\frac{\overline{7}}{3})$ reflections. These two reflections are the $(3,3,2)-(\frac{1}{3},\frac{1}{3},\frac{1}{3})$ and the $(3,0,-2)-(\frac{1}{3},\frac{1}{3},\frac{1}{3})$ reflections respectively, so the first one provides us with access to the $E_1E_2$ terms and the second main magnetic reflection allows us to probe the $E_1E_1$ and $E_2E_2$ contributions.

The main magnetic reflections (e.g. $(\frac{8}{3},\frac{\overline{1}}{3},\frac{\overline{7}}{3})$) have a complex energy spectra that has been discussed in ref.\cite{Agrestini08}. The new $(\frac{8}{3},\frac{8}{3},\frac{5}{3})$ reflection exhibits a very simple energy spectra that can be described using a single oscillator localized in the pre-edge region, which points to the interference term ($E_1E_2$) being well localized in energy in the pre-edge region. This is in good agreement with the fact that the $p$ states will be not accessible through these higher rank processes. The reduced  intensity ($\simeq \frac{1}{10}$) of the $(\frac{8}{3},\frac{8}{3},\frac{5}{3})$ reflection compared to the $(\frac{8}{3},\frac{\overline{1}}{3},\frac{\overline{7}}{3})$ reflection confirms that the former originates from a higher rank process.

\begin{figure}
\includegraphics[width=\columnwidth]{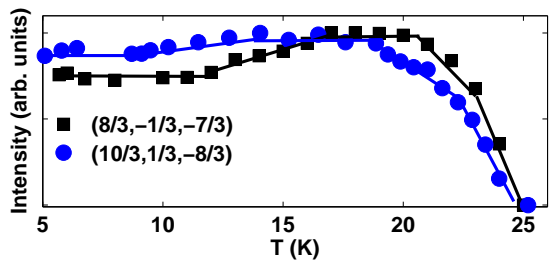}
\vspace{-0.7cm}
\caption{\label{fig:Tdep} (Color on line) Comparison of the temperature dependence of two reflections due to different scattering processes.  The lines are a guide to the eye. While the general behavior is similar, the intensity of the $E_1E_2$ reflection (circles) increases more slowly immediately below the ordering temperature than the signal due to $E_1E_1$ (squares).}
\end{figure}

\begin{figure}
\includegraphics[width=\columnwidth]{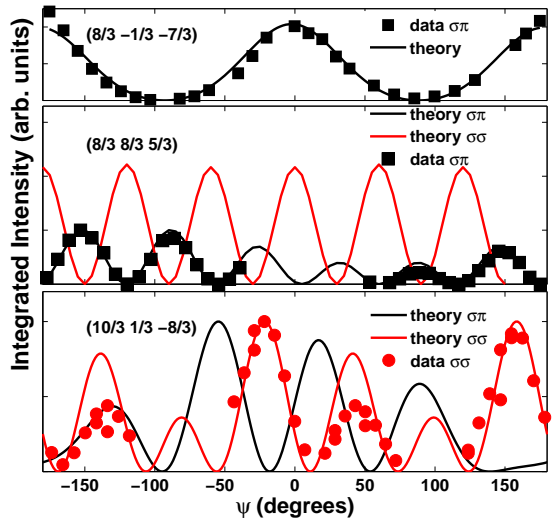}
\vspace{-0.7 cm}
\caption{\label{fig:azidep017} (Color on line) Theoretical and experimental azimuthal dependence of three magnetic reflections at the Co K-absorption edge. In the first panel an $E_1E_1$ reflection is reported, whereas the central and bottom panels show reflections resulting from the $E_1E_2$ interference together with the model predictions. The amplitude is the only free parameter in the fit.}
\end{figure}

Fig. \ref{fig:Tdep} shows the temperature dependence of the  main magnetic reflection $(\frac{8}{3},\frac{\overline{1}}{3},\frac{\overline{7}}{3})$ and of the $E_1E_2$ reflection $(\frac{10}{3} ,\frac{1}{3} , \frac{\overline{8}}{3})$.  It can be seen that the reflections not only appear at the same magnetic propagation vector as the main magnetic reflections, but that it also exhibits a very similar temperature dependence. These two facts suggest that the two families of reflections share the same physical origin. 
It is worth noting that due to the sensitivity of anomalous diffraction to small local structural changes we cannot completely exclude the possibility that an atomic displacement may follow the magnetic ordering. However, no displacements have been reported in this compound. We therefore assume that only the time odd tensors can contribute to the two family of reflections.     
The azimuthal dependence of the reflections reported in Fig. \ref{fig:azidep017} confirms that the 
$(\frac{8}{3},\frac{8}{3},\frac{5}{3})$ and the $(\frac{10}{3} ,\frac{1}{3} , \frac{\overline{8}}{3} )$ reflections are due to higher rank tensors. In fact the $(\frac{8}{3} , \frac{\overline{1}}{3} , \frac{\overline{7}}{3} )$ reflection 
due to the $E_1E_1$ process has an almost perfect 2-fold symmetry, whereas the $(\frac{8}{3},\frac{8}{3}, \frac{5}{3})$ signal, which is  $\approx 10$ degrees from the 3-fold (111) axis has 6 peaks with different intensities and 
with zeros values that are not exactly 60 degrees apart. 
The term $\tilde{F}^{3}_3 - \tilde{F}^{3}_{-3}$ has a 3-fold periodicity with respect to a rotation about
the quantization axis, whereas 
$\tilde{F}^{1}_0$ and $\tilde{F}^{3}_0$ are constant therefore they cannot reproduce the symmetry of the signal. 
Hence we will neglect these contributions in the data analysis. 

To calculate the behavior of the reflections in the reference frame of $T^p_q$ we apply a sequence of coordinate rotations ~\cite{Lovesey05} to the tensorial structural factor. This operation, together with the actual calculations of the $T^p_q$ in terms of the initial and final polarization and wave vectors, produces the following expressions for the 
$(\frac{8}{3},\frac{8}{3},\frac{5}{3})$ signal

\begin{eqnarray}
\sigma \sigma  =  a_5 T^3_3 \left[ a_6 +a_7 c_{2 \psi} \right] c_{\theta} s_{\psi} 
\end{eqnarray}
\begin{eqnarray}
\sigma \pi  =  a_1 T^3_3 c_{\psi} c_{\theta} \left[ a_2 c_{\theta} s_{\psi} +(a_3+a_4 c_{2 \psi}) s_{\theta} \right]
\end{eqnarray}
 
\noindent where $a_i$ are complex numerical coefficients   defined by the direction in the space of the reflection,  $c_x(s_x)$ are shorthand for $\cos{x}(\sin{x})$, $\theta$ is the Bragg angle and $\psi$ is the azimuthal angle. The variable $T^3_3(\omega)$ embeds the resonant denominator and the radial part of the integrals between the ground state and the intermediate state in the resonant process. A similar expression is obtained in the case of the $(\frac{10}{3} , \frac{1}{3} , \frac{\overline{8}}{3})$ reflection. As the system is not magneto-electric, in an absorption experiment it would not be possible to observe a tensorial contribution coming from the $E_1E_2$ interference. However, in a resonant diffraction experiment, we need to look at the local symmetry to establish which tensors can contribute to the scattering. The possibility of observing an $E_1E_2$ interference term in a globally centrosymmetric system has been demonstrated in the case of V$_2$O$_3$ in the time even case, where the absence of an inversion symmetry at the V position is responsible for the appearance of such a term. In \caco\ at the CoII site, both the magnetic {\sl 32'} and the non-magnetic point groups {\sl 32} allow a magneto-electric tensor to occur. A useful classification of the electromagnetic terms occurring in the RXS scattering based exclusively on the linear magneto-electric effect has been given in Refs.~\onlinecite{DiMatteo05, Marri04}. According to this scheme, the dominant term contributing to the scattering is a polar toroidal octupole (anapole). A contribution from a polar toroidal dipole ($T^3_0$) is also allowed.

The possibility of observing a polar toroidal octupole (anapole) has been widely discussed theoretically, but to date very few experimental observations of this quantity are available. V$_2$O$_3$ is certainly the most well known case \cite{Paolasini01} but in this instance the reflection structure factor allows an $E_2E_2$ term (the magnetic octupole) to occur together with the $E_1E_2$ terms \cite{DiMatteo05}. 
More recently, a possible $E_1E_2$ signal has been reported in the ferroelectric phase of TbMnO$_3$~\cite{Mannix07}. Here, both the $E_1E_2$ and the $E_1E_1$ contributions are symmetry allowed, but occur at different energies. So the tail of the $E_1E_1$, that is much stronger in intensity, can give a significant contribution at the lower energies where the $E_1E_2$ resonates.  

In \caco\ the structure factor allows one to separate completely the parity odd from the parity even events. The importance of the $E_1E_2$ interference in the analysis of RXS \cite{Marri04} and of the anapole in the electromagnetic theory\cite{Marinov07} as well as in nuclear and particle physics has been underlined in a number of experimental and theoretical papers. Ab-initio simulations have recently revealed the existence of toroidal polarizations in nanodiscs and nanorods\cite{Naumov04} and their use was suggested for data storage. Usually the existence of such effects is masked in solid state physics by other effects. In this respect RXS offers an almost unique possibility of accessing such effects.  Other sources of scattering (e.g. non-magnetic $E_1E_2$ interference) cannot be ruled out by the geometrical dependence of the RXS signal alone. Nevertheless, the absence of any structural change and the thermal behavior of the signal strongly suggests that a form of anapolar ordering develops in the low temperature phase of \caco.  In conclusion, we demonstrate once again that RXS is a powerful technique for accessing subtle features in the charge distribution in complex materials and that there is an urgent need to develop further the theoretical interpretation of the data.      
     
The authors acknowledge useful discussions with S. W. Lovesey, V. E. Dmitrienko, and S. Di Matteo.
This work was supported by a grant from the EPSRC, UK (EP/C000757/1). We acknowledge the European Synchrotron Radiation Facility for provision of beam time.


\begin{thebibliography}{100}

\bibitem{Agrestini08} S. Agrestini, C. Mazzoli, A. Bombardi, and M. R. Lees Phys. Rev. B \textbf{77}, 140403(R) (2008).
\bibitem{Fjellvag96} H. Fjellv{\aa}g, E. Gulbrandsen, S. Aasland, A. Olsen, and B. C. Hauback, J. Solid State Chem. \textbf{124}, 190 (1996).
\bibitem{Sampa04} E. V. Sampathkumaran,  N. Fujiwara, S. Rayaprol, P. K. Madhu and Y. Uwatoko, Phys. Rev. B \textbf{70}, 014437 (2004).
\bibitem{Burnus06} T. Burnus, Z. Hu, M. W. Haverkort, J. C. Cezar, D. Flahaut, V. Hardy,  A. Maignan,  N. B. Brookes,  A. Tanaka, H. H. Hsieh, H.-J. Lin, C. T. Chen, and L. H. Tjeng, Phys Rev. B \textbf{74}, 245111 (2006).
\bibitem{Kageyama97} H. Kageyama, K. Yoshimura, K. Kosuge, M. Azuma, M. Takano, H. Mitamura, and T. Goto, J. Phys. Soc. Jpn. \textbf{66}, 3996 (1997).
\bibitem{Lovesey05} S. W. Lovesey, E. Balcar, K. S. Knight, and J. F. Rodriguez, Phys. Rep. \textbf{411}, 233 (2005).
\bibitem{DiMatteo03} S. Di Matteo, Y. Joly, A. Bombardi , L. Paolasini, F. de Bergevin, and C. R. Natoli, Phys. Rev. Lett. \textbf{91}, 257402 (2003).
\bibitem{Carra94} P. Carra and  B. T. Thole, Rev. Mod. Phys. \textbf{66}, 1509 (1994).
\bibitem{Dmitrienko} V. E. Dmitrienko, Acta Cryst.  A \textbf{39}, 29 (1983).
\bibitem{Paolasini01} L. Paolasini, S. Di Matteo, C. Vettier, F. de Bergevin, A. Sollier, W. Neubeck, F. Yakhou, P. A. Metcalf, and J. M. Honig, J. Electron Spectrosc. Relat. Phenom. \textbf{120}, 1 (2001).
\bibitem{Paolasini07} L. Paolasini, C. Detlefs, C. Mazzoli, S. Wilkins, P. P. Deen, A. Bombardi, N. Kernavanois, F. de Bergevin, F. Yakhou, J. P. Valade, I. Breslavetz, A. Fondacaro, G. Pepellin, and P. Bernard, J. of Synch. Rad. \textbf{14}, 301 (2007).
\bibitem{DiMatteo05} S. Di Matteo, Y. Joly, and R. Natoli, Phys. Rev. B \textbf{72}, 144406 (2005).
\bibitem{Marri04} I. Marri and P. Carra, Phys. Rev. B \textbf{69}, 113101 (2004). 
\bibitem{Mannix07} D. Mannix, D. F. McMorrow, R.A. Ewings, A. T. Boothroyd, D. Prabhakaran, Y. Joly, B. Janousova, C. Mazzoli, L. Paolasini, and S. B. Wilkins, Phys. Rev. B \textbf{76}, 184420 (2007).
\bibitem{Marinov07} K. Marinov, A. D. Boardman, V. A. Fedotov, N. Zheludev, New J. Phys. \textbf{9} 324 (2007). 
\bibitem{Naumov04} I. I. Naumov, L. Bellaiche, and H. Fu, Nature(London) \textbf{432}, 737 (2004).

\end{thebibliography}
\end{document}